\documentclass[conference]{IEEEtran}
\IEEEoverridecommandlockouts
\usepackage{cite}
\usepackage{amsmath,amssymb,amsfonts}
\usepackage{algorithmic}
\usepackage{graphicx}
\usepackage{textcomp}
\usepackage{xcolor}
\usepackage[caption=false,font=footnotesize]{subfig}

\begin{document}

\title{ChannelAgent-Empowered Electromagnetic Space World Model: A Case Study on Agent-Driven Channel Generation for 6G AI-Native Air Interface\\

\thanks{This work is supported by the National Natural Science Foundation of China (No. 62401084, No. 62525101 and No. 62401068), and the Beijing University of Posts and Telecommunications - China Mobile Communications Group Co., Ltd. Joint Institute.}
}

\author{\IEEEauthorblockN{Mingyue Li$^{1}$, Li Yu$^{1}$, Yuxiang Zhang$^{1}$, Heng Wang$^{1}$, Jianhua Zhang$^{1}$, Ping Zhang$^{1}$, Guangyi Liu$^{2}$}

\IEEEauthorblockA{$^{1}$\textit{State Key Laboratory of Networking and Switching Technology,} \\
\textit{Beijing University of Posts and Telecommunications, Beijing 100876, China}\\
$^{2}$\textit{China Mobile Research Institute, China}\\
Emails: \{mingyueli, li.yu, zhangyx, hengwang, jhzhang, pzhang\}@bupt.edu.cn, liuguangyi@chinamobile.com}
}
\maketitle

\begin{abstract}

As sixth-generation (6G) wireless networks evolve toward increasingly heterogeneous scenarios, tasks, and service requirements, conventional artificial intelligence (AI) models remain limited in task-aware decision-making and autonomous adaptation. To address this issue, this paper first proposes a ChannelAgent-empowered electromagnetic space world model, in which wireless intelligence is organized into a closed-loop process consisting of multi-modal sensing, ChannelAgent as the intelligent core, and execution with feedback update. As a case study, agent-driven channel generation is instantiated through path loss prediction. Specifically, a task-oriented intelligent feature selection mechanism is designed by integrating reinforcement-learning-inspired policy adaptation with evolutionary search, enabling the agent to iteratively derive compact and task-suitable feature subsets according to the current scenario and performance feedback. Simulation results demonstrate superior performance in both single-scenario and multi-scenario tasks, highlighting the potential of the proposed model for autonomous, adaptive, task-oriented, and closed-loop wireless intelligence.

\end{abstract}


\vspace{-2pt}
\section{Introduction}
\vspace{-2pt}
As sixth-generation (6G) wireless systems continue to evolve, future networks are expected to operate in increasingly diverse and dynamic scenarios. Emerging paradigms such as integrated sensing and communication, artificial intelligence (AI)-native air interface, and ubiquitous connectivity are continuously expanding the functional scope of wireless systems \cite{ref1, ref2, ref3}. Meanwhile, multi-band coexistence, complex deployment environments, and diverse key performance indicator (KPI) requirements pose heterogeneous challenges in scenarios, tasks, and performance objectives. Linking electromagnetic (EM) space with communication performance, wireless channels are closely coupled with propagation conditions and service requirements. Therefore, proactive sensing, understanding, and adaptation of channel states and communication tasks become critical for future wireless systems.

To meet this demand, recent advances in AI have introduced powerful data-driven tools into wireless communications. From task-specific neural network predictors \cite{ref4} to emerging large models \cite{ref5}, AI-based methods have demonstrated promising performance in a wide range of communication tasks, including channel prediction \cite{ref6}, beam management \cite{ref7}, radio map construction \cite{ref8}, and resource allocation \cite{ref10}. However, most existing approaches still function primarily as algorithmic tools under predefined task pipelines: they mainly learn passive input-output mappings for specific tasks, lacking the autonomous decision-making capability to proactively adapt to changing environments and dynamic service requirements. 

This limitation suggests that future wireless systems should move beyond isolated predictive models toward an autonomous and closed-loop wireless intelligence paradigm  \cite{ref11}. In this context, AI agents offer a promising direction. An AI agent can be regarded as an autonomous computational entity that perceives its environment, performs reasoning and decision-making, and continuously interacts with the environment through actions to accomplish task objectives \cite{ref12, ref13}. Compared with conventional AI models that mainly provide predictive capabilities, an agent further emphasizes task interpretation, decision-making, tool orchestration, and feedback-driven adaptation, thereby transforming model capability into task-oriented action capability. This property is particularly important for future wireless systems, where communication decisions must adapt to task objectives, environmental dynamics, and propagation conditions.

Motivated by this insight, this paper introduces the concept of AI agents into the channel domain and first proposes a ChannelAgent-empowered EM space world model for multi-dimensional communication tasks in the wireless EM world. Rather than treating communication tasks as predefined prediction problems under fixed input-output mappings, the proposed model is designed as a closed-loop architecture consisting of multi-modal sensing, ChannelAgent with intelligent brain core and knowledge and tool support, and execution with feedback update, moving wireless systems beyond passive algorithm invocation toward a more intelligent paradigm with autonomous adaptation, independent decision-making, task-oriented operation, and closed-loop feedback update. In this sense, the proposed model provides a potential agent-oriented paradigm for future 6G AI-native air interface.

Furthermore, this paper develops agent-driven channel generation as a case study to demonstrate the capability of the proposed model for task-oriented decision under different scenarios. Specifically, an intelligent feature selection mechanism is designed by integrating reinforcement-learning-inspired policy adaptation with evolutionary search, enabling the agent to iteratively derive a more compact and task-suitable feature subset according to the current scenario and performance feedback. Using path loss prediction as a representative task, simulation results show that the proposed method achieves superior performance in both single-scenario and multi-scenario tasks, providing an initial task-level validation of the proposed model.

The remainder of this paper is organized as follows. Section \ref{sec2} presents the proposed model. Section \ref{sec3} introduces the case study on agent-driven channel generation. Section \ref{sec4} provides simulations and performance analysis. Finally, Section \ref{sec5} gives the conclusion of this paper.

\begin{figure*}[htbp]
    \centering 
    \captionsetup{skip=-4pt}
    \includegraphics[width=0.9\textwidth]{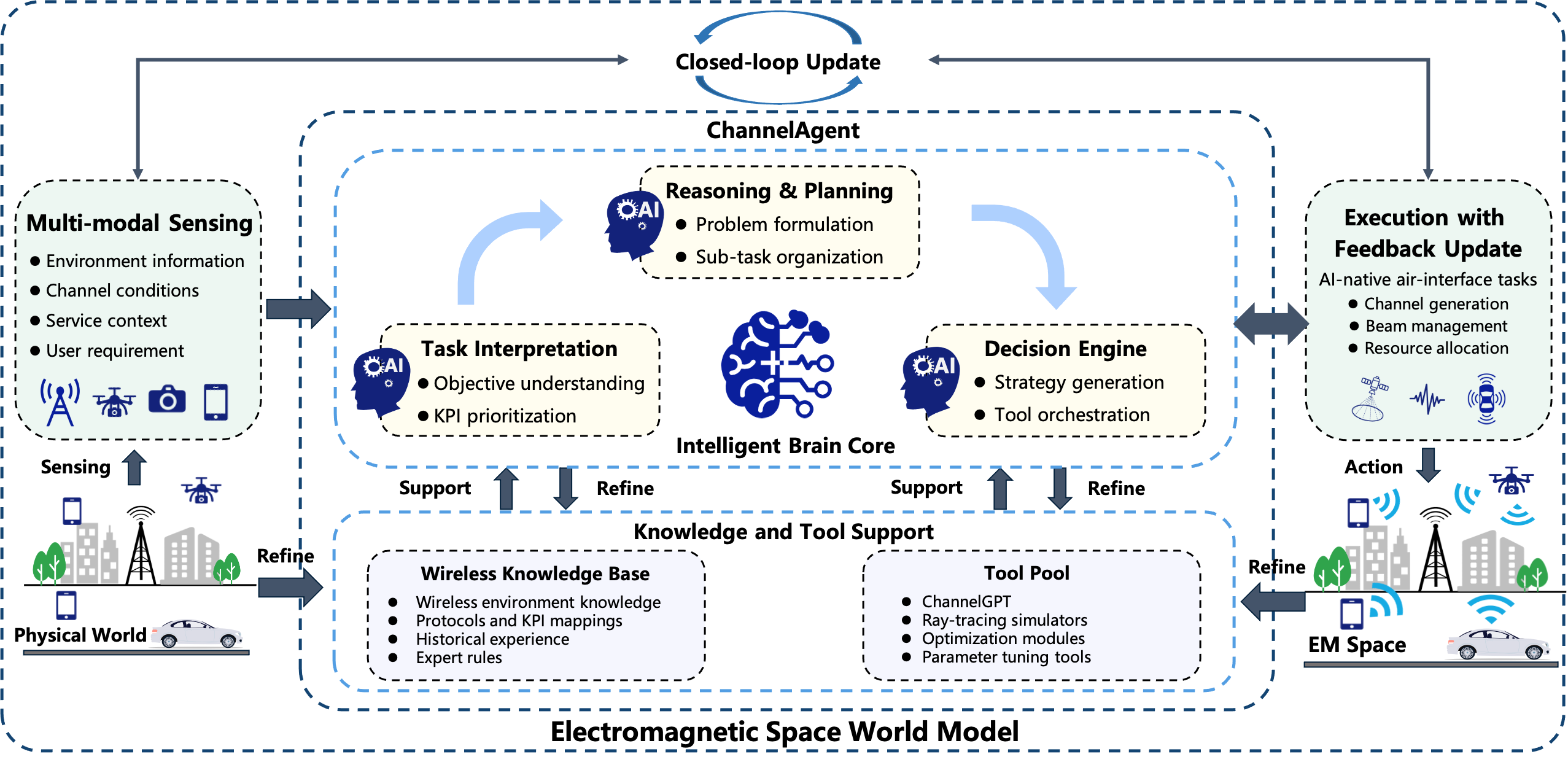}
    \caption{Architecture of the proposed ChannelAgent-empowered electromagnetic space world model for future 6G AI-native air interface.}
    \label{fig_ChannelAgent}
    \vspace{-10pt}
\end{figure*}

\vspace{-5pt}
\section{The ChannelAgent-Empowered Electromagnetic Space World Model}\label{sec2}
\vspace{-5pt}
This section presents the proposed ChannelAgent-empowered EM space world model for multi-dimensional communication tasks in the wireless EM world. As illustrated in Fig.~\ref{fig_ChannelAgent}, the proposed model consists of three components: multi-modal sensing, ChannelAgent, and execution with feedback update. Within this model, ChannelAgent integrates the intelligent brain core with knowledge and tool support. Rather than following a passive fitting pipeline for predefined tasks, the proposed model forms a task-oriented closed-loop architecture that enables wireless systems to sense the wireless EM space, generate decisions, execute actions, and refine subsequent behavior through feedback.

\vspace{-5pt}
\subsection{Multi-modal Sensing}
\vspace{-3pt}
Serving as the sensory interface of the proposed model, multi-modal sensing is responsible for acquiring external state information relevant to the current communication task and forming an initial representation of the wireless EM space. Unlike conventional systems that mainly rely on local link measurements, this component supports a multi-dimensional understanding of the wireless EM space, jointly characterizing channel conditions, environmental context, user requirements, and service-related constraints, thereby establishing a more holistic sensing of the current scenario. Since communication tasks and their KPI priorities may vary across scenarios, the sensing information should jointly characterize environmental conditions and task constraints. Accordingly, multi-modal sensing provides the input basis for subsequent processes.

\subsection{Intelligent Brain Core}
\vspace{-3pt}
As the cognitive center within ChannelAgent, the intelligent brain core is responsible for transforming sensing inputs into executable communication decisions for the current task. Rather than functioning as a single, black-box model, it undertakes the core functions of task interpretation, problem formulation, and decision generation, enabling the system to infer task objectives from sensing inputs and organize subsequent actions accordingly.

Specifically, this core contains three closely coupled stages. During task interpretation, the system identifies the current task objective by jointly considering the scenario state, service requirements, and performance constraints, and determines the KPIs to be prioritized. Subsequently, in reasoning and planning, it formulates the problem according to the environment state and task requirements, and converts the overall objective into an executable task chain. Finally, within the decision engine, the system determines the task strategy and identifies the tools required for subsequent execution.

Through this process, ChannelAgent does not simply execute a fixed algorithm for a given input, but autonomously performs understanding, problem formulation, and decision-making around the task objective. For example, when a vehicle turns at an intersection, the system can dynamically determine whether propagation state re-evaluation, beam reselection, or channel prediction update is required according to environmental changes, link conditions, and service demands.

\subsection{Knowledge and Tool Support}
\vspace{-3pt}
Knowledge and tool support provides the essential prior grounding and execution support for ChannelAgent within the proposed model. It complements online sensing and reasoning by supplying both reusable domain knowledge and callable functional modules.

The wireless knowledge base mainly stores communication-related prior knowledge, such as wireless environment knowledge (WEK)\cite{sim-3}, protocol rules, mappings between KPIs and task objectives, historical decision experience, and expert rules. Such knowledge provides important references for task interpretation and reasoning, allowing the core decision process to go beyond instantaneous inputs and perform more informed problem formulation and objective decomposition.

The tool pool, on the other hand, provides the executable modules required for task realization, such as neural prediction backbones, ray-tracing simulators, optimization modules, and parameter adjustment tools. These tools do not operate independently. Instead, they are invoked on demand under the coordination of the decision engine to support the execution of different communication tasks.

\subsection{Execution with Feedback Update}
\vspace{-3pt}
Execution with feedback update is responsible for implementing the generated decisions in the wireless EM space and refining the proposed model according to execution outcomes. The selected strategy is carried out through the required tools to accomplish the corresponding communication task, thereby translating the outputs of ChannelAgent into effective actions in the wireless EM space. The resulting performance is then evaluated and fed back to ChannelAgent while also being used to refine the knowledge and tool support. Through this process, the proposed model continuously improves future decisions and completes the closed-loop interaction with the wireless EM space.

Overall, the proposed model establishes a closed-loop intelligent architecture for the wireless EM world, featuring autonomous adaptation, independent decision-making, task-oriented operation, and closed-loop feedback update. It has the potential to drive wireless systems to move from passive model execution toward a proactive paradigm of autonomous understanding, decision-making, and optimization around current task objectives.

\begin{figure*}[htbp]
\centering    
\captionsetup{skip=-2pt}  
\includegraphics[width=0.9\textwidth]{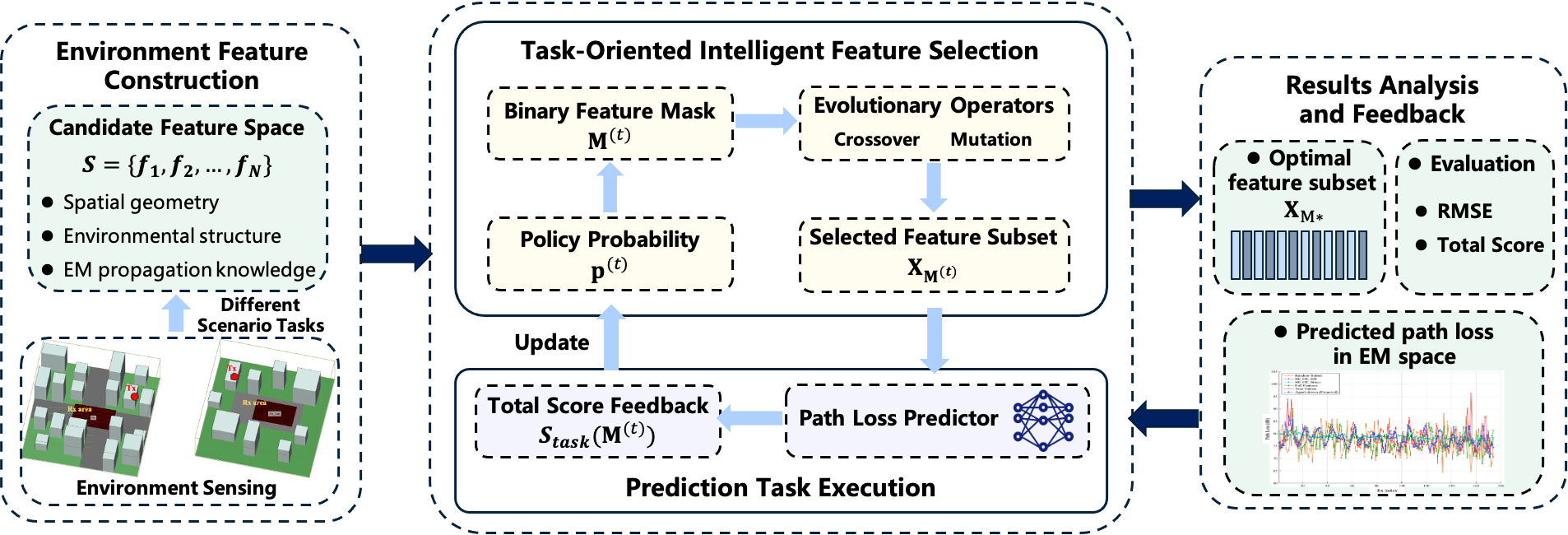}
\caption{Workflow of the proposed agent-driven channel generation case study.}
\label{fig_PL}
\vspace{-10pt}    
\end{figure*}

\section{Case Study: Agent-driven Channel Generation}\label{sec3} 

This section presents agent-driven channel generation as a case study of the proposed model. The process is instantiated through path loss prediction, a representative channel-related task strongly affected by environmental geometry and propagation conditions. Unlike conventional methods with fixed feature inputs, the proposed workflow enables task-oriented intelligent feature selection through a hybrid mechanism that combines reinforcement-learning-inspired policy adaptation with evolutionary search, thereby automatically deriving a compact and task-suitable feature subset for accurate path loss prediction under different scenario-dependent tasks. Specifically, as shown in Fig.~\ref{fig_PL}, the workflow consists of environment feature construction, task-oriented intelligent feature selection, prediction task execution, and result analysis and feedback.

\vspace{-5pt}
\subsection{Environment Feature Construction}
\vspace{-3pt}
The workflow begins with the construction of a candidate feature space for path loss prediction. Following environment sensing and preprocessing in the given scenario, raw data are abstracted into a set of representative environmental features to provide the basis for subsequent decision-making. As summarized in Table \ref{tab:feature_list}, ten environmental features are constructed to capture the principal factors governing path loss from three complementary perspectives: spatial geometry, environmental structure, and EM propagation knowledge \cite{sim-1,sim-2,sim-3}. Accordingly, the candidate feature space is defined as\begin{equation}
{\mathcal{S}} = \{f_1, f_2, \dots, f_N\},
\end{equation} where $f_i$ denotes the $i$-th candidate environment feature and $N$ is the total number of available features. Instead of fixing the model input a priori, the proposed method retains all features in $\mathcal{S}$ as candidate variables for subsequent task-oriented intelligent feature selection.

\vspace{-5pt}
\begin{table}[htbp]
\centering
\vspace{-8pt}     
\captionsetup{skip=3pt}  
\caption{Candidate environment features used in the case study}
\label{tab:feature_list}
\begin{tabular}{|c|c|p{3.8cm}|}
\hline
\textbf{Category} & \textbf{Symbol} & \textbf{Meaning} \\
\hline
Geometry (1) & $D_{txrx}$ & Tx--Rx distance \\
\hline
Geometry (2) & $H_{txrx}$ & Tx--Rx height difference \\
\hline
Geometry (3) & $H_{ts\_avg}$ & Tx--scatterer height difference \\
\hline
Geometry (4) & $D_{ts\_avg}$ & Tx--scatterer distance \\
\hline
Geometry (5) & $D_{rs\_mean}$ & Rx--scatterer distance \\
\hline
Structure (6) & $V_{eff\_mean}$ & Mean scatterer volume \\
\hline
Structure (7) & $Min\_D_{dev\_eff}$ & Minimum offset to Tx--Rx line \\
\hline
Structure (8) & $Blockage_{eff}$ & Blockage feature \\
\hline
Knowledge (9) & $Ref\_WEK$ & Reflection contribution \\
\hline
Knowledge (10) & $Dif\_WEK$ & Diffraction contribution \\
\hline
\end{tabular}
\vspace{-10pt}
\end{table}

\vspace{-2pt}
\subsection{Task-Oriented Intelligent Feature Selection}
\vspace{-2pt}
Given the constructed candidate feature space, the workflow performs task-oriented intelligent feature selection for path loss prediction in the current scenario. The objective is not to retain as many input features as possible, but to identify a compact and task-suitable subset that better supports the downstream prediction task. To this end, each candidate solution is represented by a binary feature mask: \begin{equation}
\mathbf{M}^{(t)} = \left[ m_1^{(t)}, m_2^{(t)}, \dots, m_N^{(t)} \right] \in \{0, 1\}^N,
\label{eq:mask_vector}
\end{equation} where ${m_i}^{(t)} = 1$ indicates that the $i$-th feature is selected at generation $t$, and ${m_i}^{(t)} = 0$ otherwise.

The proposed workflow employs a hybrid mechanism that integrates reinforcement-learning-inspired policy adaptation with evolutionary search, enabling the generated feature subsets to be iteratively refined to match the current scenario and task objective. Specifically, the feature-selection policy is parameterized by a probability vector:\begin{equation}
\mathbf{p}^{(t)} = [p_1^{(t)}, p_2^{(t)}, \dots, p_N^{(t)}],\label{p}
\end{equation} where $p_i^{(t)}$ denotes the probability of selecting the $i$-th feature at generation $t$. Under this policy, each candidate mask is sampled according to:\begin{equation}
m_i^{(t)} \sim \text{Bernoulli}\left( p_i^{(t)} \right), \quad i = 1, 2, \dots, N.
\label{eq:bernoulli}
\end{equation}

To enhance exploration over the combinatorial feature space, evolutionary operators such as crossover and mutation are further applied to the sampled masks. The generated feature subsets are subsequently evaluated through the downstream prediction task. To align the decision process with the task objective, the performance is converted into a total score that jointly considers prediction error, variation-trend consistency, and feature compactness:
\begin{equation}
\label{eq:total_score}
\mathcal{S}_{\text{task}}(\mathbf{M}) =
-\left(
\mathrm{RMSE}(\mathbf{M})
+ \lambda_{c} \mathcal{E}_{\text{c}}(\mathbf{M})
+ \frac{\lambda_n}{N}\|\mathbf{M}\|_0
\right),
\end{equation}
where $\mathrm{RMSE}(\mathbf{M})$ denotes the validation root mean square error under feature mask $\mathbf{M}$, 
$\mathcal{E}_{\text{c}}(\mathbf{M})$ denotes the trend-consistency error between the predicted and ground-truth path loss sequences, 
encouraging the selected feature subset to better preserve local path loss variations induced by blockage and shadowing. 
$\|\mathbf{M}\|_0$ denotes the number of selected features, and $\lambda_{c},\lambda_n$ are weighting coefficients. 
A larger $\mathcal{S}_{\text{task}}(\mathbf{M})$ indicates better overall performance.

Based on the obtained total score, the top-$K$ candidate masks are selected as elite individuals. Their average feature distribution, denoted by $\bar{\mathbf{M}}^{(t)}_{\mathrm{elite}} = \frac{1}{K} \sum_{k=1}^{K} \mathbf{M}^{(t)}_{\mathrm{elite},k}$ is then utilized to update the global policy via a moving average:\begin{equation}
\mathbf{p}^{(t+1)} = (1 - \eta) \mathbf{p}^{(t)} + \eta \, \bar{\mathbf{M}}^{(t)}_{\mathrm{elite}},
\end{equation} where $\eta$ is the update rate. Through this score-guided iterative process, the agent progressively refines its feature preference toward subsets that are better aligned with the current scenario and task objective.

\subsection{Prediction Task Execution}
\vspace{-2pt}

After the task-oriented intelligent feature selection, the optimal feature mask is determined as\begin{equation}
\mathbf{M}^{\star} = \arg\max_{\mathbf{M}} \mathcal{S}_{\mathrm{task}}(\mathbf{M}),
\end{equation} which corresponds to the feature subset with the highest total score for the current scenario. Based on $\mathbf{M}^{\star}$, the associated input features are extracted and fed into the path loss predictor to perform the final prediction task. Denoting the selected feature subset by $\mathbf{X}_{\mathbf{M}^{\star}}$, the prediction process can be written as
\begin{equation}
\hat{\mathbf{y}} = f_{\theta}\!\left(\mathbf{X}_{\mathbf{M}^{\star}}\right),
\end{equation} where $f_{\theta}(\cdot)$ denotes the path loss predictor with parameter set $\theta$, and $\hat{\mathbf{y}}$ is the predicted path loss. In this way, the feature subset obtained by the agent is finally translated into executable path loss prediction for the current scenario.

\subsection{Results Analysis and Feedback}
\vspace{-2pt}

For each scenario, the proposed workflow yields a task-specific decision outcome composed of the optimal feature mask $\mathbf{M}^{\star}$, the corresponding prediction result $\hat{\mathbf{y}}$, and its performance evaluation. Therefore, the agent-driven method provides not only the final path loss prediction, but also an interpretable feature-level decision associated with the current task. The evaluation result further serves as feedback to the decision module, thereby closing the loop between feature selection and prediction performance.

Since the relevance of candidate features varies with the underlying propagation environment, the optimal feature subset is inherently scenario-dependent. By comparing $\mathbf{M}_{s}^{\star}$ across different scenarios, the proposed workflow can reveal which environmental features are more critical under different propagation conditions. Therefore, when the scenario changes, the agent can dynamically update its feature preference accordingly and generate a new task-matched subset instead of using a fixed input configuration.

\vspace{-2pt}
\section{Simulation Setup and Results}\label{sec4} 
\vspace{-2pt}
\subsection{Simulation Setup}
\begin{figure}[htbp]
    \centering
    \vspace{-10pt}
    \begin{minipage}[b]{0.45\linewidth}
        \centering
        \subfloat[Task 1: An intersection scenario.\label{fig_sce1}] {
            \resizebox{\linewidth}{3cm}{\includegraphics{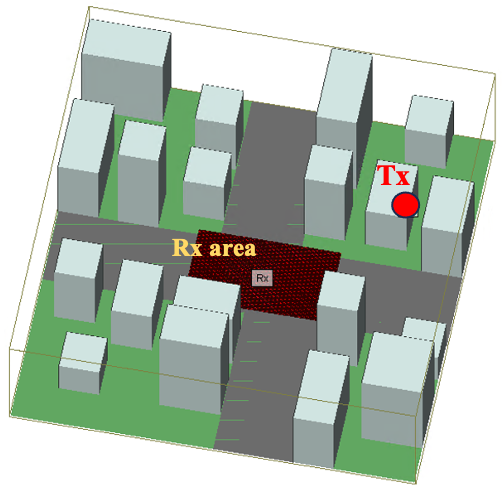}} 
        }
    \end{minipage}
    \hspace{0.02\linewidth}
    \begin{minipage}[b]{0.45\linewidth}
        \centering
        \subfloat[Task 2: A city square scenario.\label{fig_sce2}]{
            \resizebox{\linewidth}{3cm}{\includegraphics{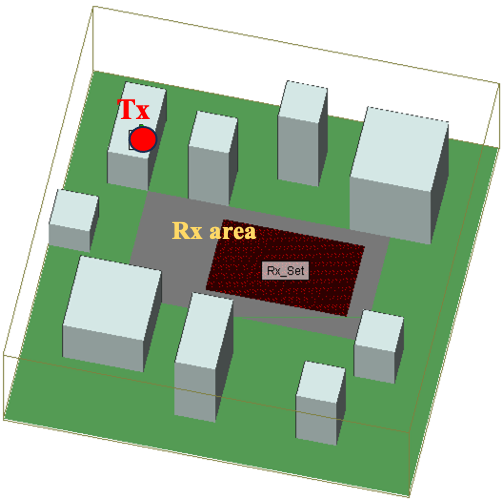}}  
        }
    \end{minipage}
    \caption{Simulation scenarios in the case study.}
    \label{fig_scenarios}
    \vspace{-10pt}
\end{figure}

\vspace{-2pt}
To evaluate the proposed workflow, three tasks are considered in this section: Task 1 corresponds to an intersection scenario, shown in Fig.~\ref{fig_sce1}, Task 2 corresponds to a city square scenario, shown in Fig.~\ref{fig_sce2}, and Task 3 is a multi-scenario task constructed by combining data samples from the two scenarios. Wireless InSite is employed to generate the path loss data together with the corresponding geometric scenario information, from which the ten candidate environment features in Table~\ref{tab:feature_list} are further extracted by scripts.

For all tasks, the same convolutional neural network (CNN)-based predictor is adopted as the downstream prediction backbone to ensure fair comparison across different feature-input strategies. In addition to the proposed agent-driven method, four comparison strategies are considered, including full-feature input \cite{sim-2}, a random subset with the same cardinality as the agent-selected one, a mutual-information-based geometric-structure subset (MI\_GE\_Struct), and a mutual-information-based geometric-EM subset (MI\_GE\_EM). The two mutual-information (MI)-based subsets are constructed according to feature relevance to the current prediction task.

The major implementation settings are summarized in Table \ref{tab_config}. The final performance is evaluated by the RMSE and the total score $\mathcal{S}_{\text{task}}(\mathbf{M})$ defined in \eqref{eq:total_score}, where a higher total score indicates better overall performance.

\begin{table}[htbp]
\centering
\captionsetup{skip=2pt}
\vspace{-10pt}
\caption{Major implementation settings}
\label{tab_config}
\begin{tabular}{|c|c|}
\hline
\textbf{Parameter} & \textbf{Value/Description} \\ \hline
CNN batch size & 16 \\ \hline
CNN loss function & RMSE \\ \hline
CNN learning rate & 0.00055 \\ \hline
Agent generations & 50 \\ \hline
Agent population size & 25 \\ \hline
Policy update rate $\eta$ & 0.1 \\ \hline
Total score coefficients $\lambda_{c},\lambda_n$ &  0.3, 0.3 \\ \hline
\end{tabular}
\vspace{-14pt}
\end{table}

\subsection{Performance Analysis}

\begin{figure}[htbp]
    \centering
    \captionsetup{skip=-2pt}
    \includegraphics[width=1\linewidth]{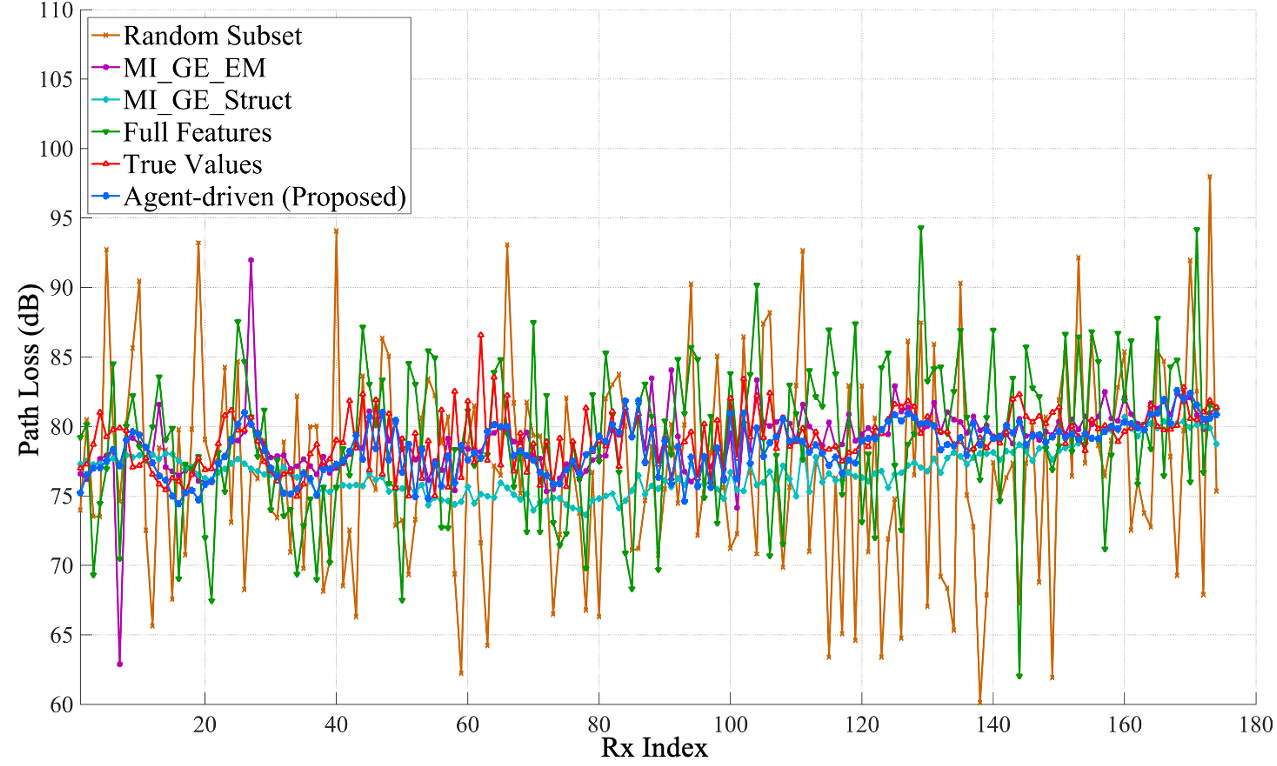}
    \caption{Path loss prediction results of the proposed method and the comparative methods in single-scenario task.}
    \label{fig_task1_2}
    \vspace{-13pt}
\end{figure}

The quantitative results are summarized in Table~\ref{tab_compare}. For the single-scenario tasks, the proposed agent-driven method achieves the best performance in both Task 1 and Task 2, reducing the RMSE from 4.823 dB to 3.736 dB and from 6.153 dB to 2.664 dB compared with the full-feature baseline, corresponding to reductions of 22.5\% and 56.7\%, respectively. As shown in Fig.~\ref{fig_task1_2}, the predicted path loss curves also better follow the ground-truth variations. These results indicate that full-feature input may introduce redundant or task-irrelevant information, while MI-based subsets, although competitive, are limited by static feature relevance and lack feedback-driven preference refinement.

\begin{figure}[htbp]
    \centering
    \captionsetup{skip=-2pt}
    \includegraphics[width=1\linewidth]{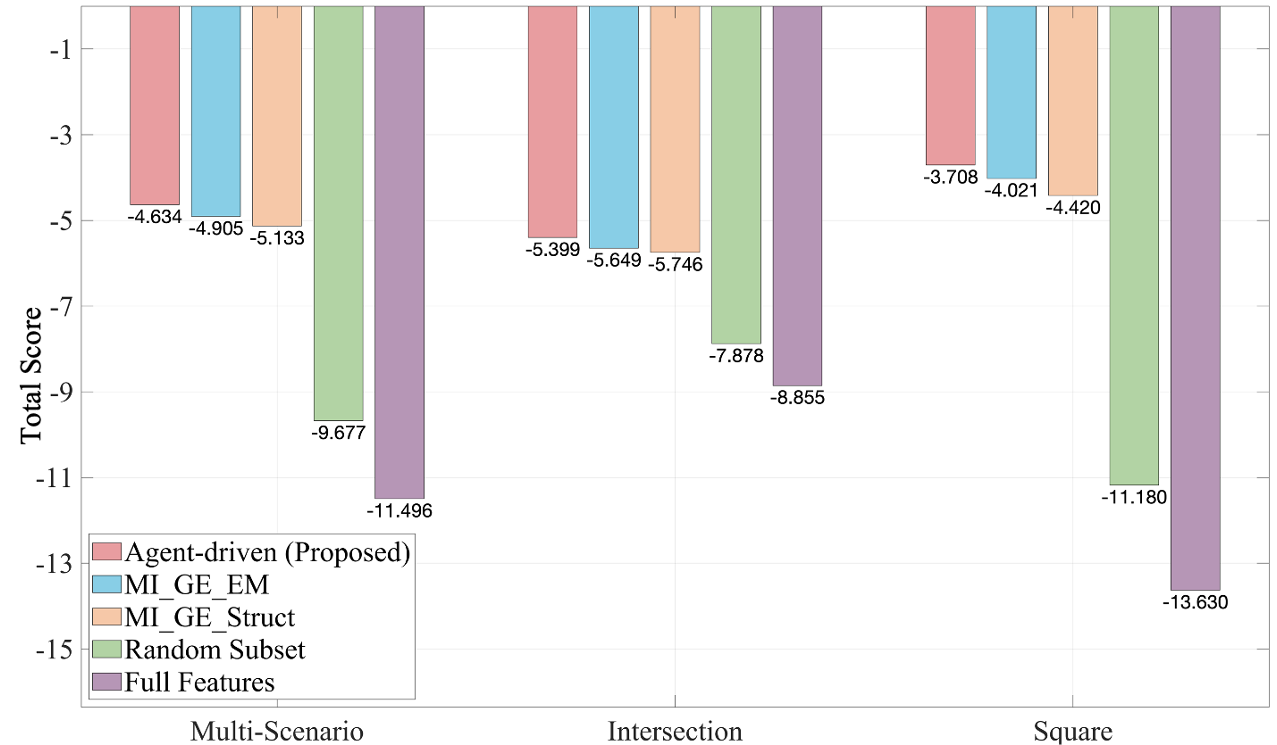}
    \caption{Generalization performance of the proposed method and the comparative methods in multi-scenario task.}
    \label{fig_task3}
    \vspace{-8pt}
\end{figure}

The advantage of the proposed method is further observed in Task 3. As shown in Fig.~\ref{fig_task3}, the proposed method achieves the highest total score on the multi-scenario task. It also maintains strong performance on the two constituent sub-scenarios, with RMSE of 3.918 dB and 2.731 dB, respectively, as shown in Table~\ref{tab_compare}. The performance gap from the corresponding single-scenario results is within 0.2 dB, indicating good cross-scenario generalization of the selected subset.

\vspace{-10pt}

\begin{table}[htbp]
\centering
\captionsetup{skip=2pt}
\vspace{-10pt}
\caption{Performance comparison under different tasks and methods}
\label{tab_compare}
\resizebox{\linewidth}{!}{
\begin{tabular}{|c|c|c|c|c|}
\hline
\textbf{Task} & \textbf{Method} & \textbf{Features} & \textbf{RMSE (dB)} & \textbf{Total score}\\
\hline
\textbf{Task 1} & \textbf{Agent-driven} & \textbf{(2,7,9,10)} & \textbf{3.736} & \textbf{-5.242}\\
\hline
Task 1 & Full-feature & All & 4.823 & -6.860\\
\hline
Task 1 & Random subset & (1,5,6,7) & 5.725 & -8.018\\
\hline
Task 1 & MI\_GE\_EM & (1,5,9,10) & 3.829 & -5.380\\
\hline
Task 1 & MI\_GE\_Struct & (1,5,7,8) & 4.174 & -5.633\\
\hline
\textbf{Task 2} & \textbf{Agent-driven} & \textbf{(3,8,9,10)} & \textbf{2.664} & \textbf{-3.646}\\
\hline
Task 2 & Full-feature & All & 6.153 & -7.465\\
\hline
Task 2 & Random subset & (1,3,6,7) & 8.750 & -11.446\\
\hline
Task 2 & MI\_GE\_EM & (1,5,9,10) & 3.083 & -4.292 \\
\hline
Task 2 & MI\_GE\_Struct & (1,5,7,8) & 3.639 & -4.599\\
\hline
\textbf{Task 3 (Multi)} & \textbf{Agent-driven} & \textbf{(3,7,9,10) }& \textbf{3.377} & \textbf{-4.634}\\
\quad -- Intersection & Agent-driven & (3,7,9,10) & 3.918 & -5.399 \\
\quad -- City Square & Agent-driven & (3,7,9,10) & 2.731 & -3.708 \\
\hline
\end{tabular}}
\vspace{-15pt}
\end{table}

\begin{figure}[htbp]
    \centering
    \captionsetup{skip=-5pt}
    \includegraphics[width=1\linewidth]{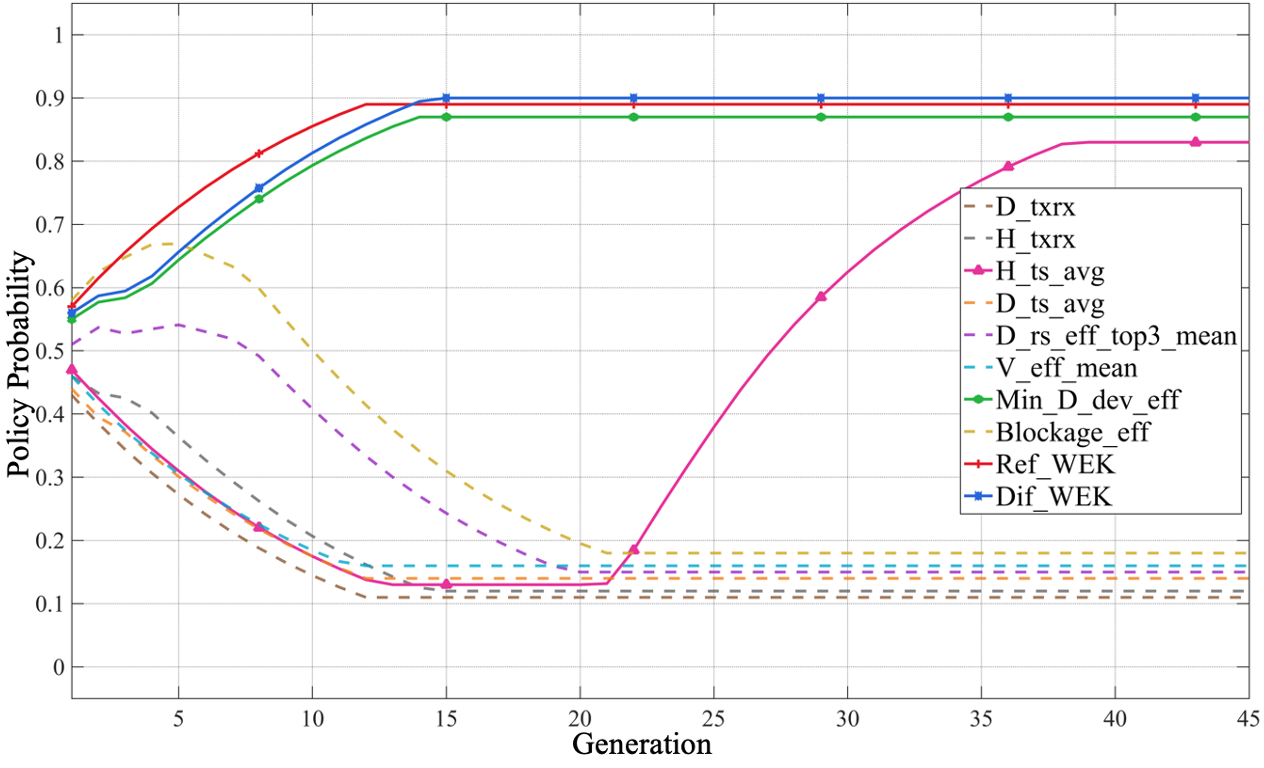}
    \caption{Evolution of the agent policy probabilities.}
    \label{fig_task3_policy}
    \vspace{-8pt}
\end{figure}

\begin{figure}[htbp]
    \centering
    \captionsetup{skip=-5pt}
    \includegraphics[width=1\linewidth]{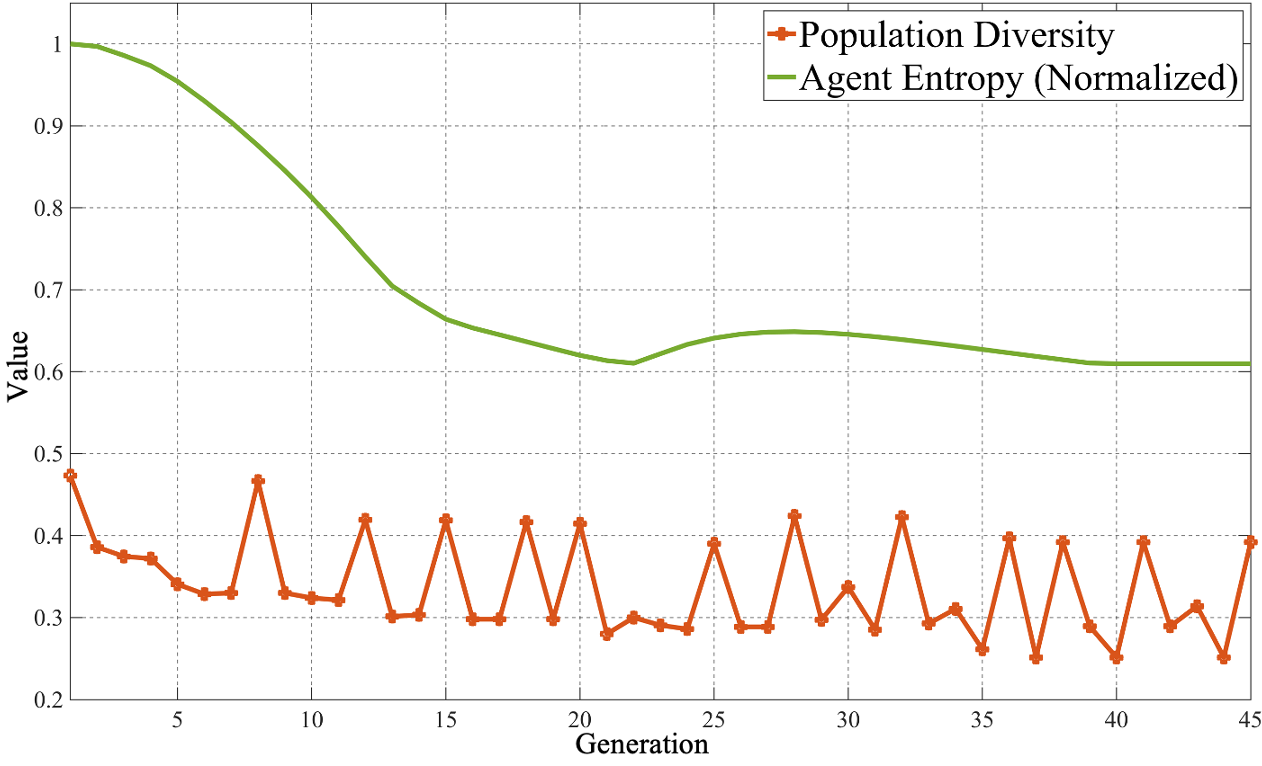}
    \caption{Evolution of normalized agent entropy and population diversity.}
    \label{fig_task3_agent}
\end{figure}

The decision evolution of the agent is further analyzed in Figs.~\ref{fig_task3_policy} and \ref{fig_task3_agent}. Since $\mathbf{p}^{(t)}$ represents the feature-selection policy defined in \eqref{p}, the trajectory of each $p_i^{(t)}$ reflects the agent's preference for the corresponding feature. To interpret this decision process, the normalized agent entropy and population diversity are considered to characterize policy determinacy and population-level exploration:
\vspace{-10pt}
\begin{equation}
\label{eq:entropy_diversity}
\begin{aligned}
\bar{H}^{(t)}
=&-\frac{1}{N\ln 2}\sum_{i=1}^{N}
\left[
p_i^{(t)}\ln p_i^{(t)}
+\left(1-p_i^{(t)}\right)\ln\left(1-p_i^{(t)}\right)
\right],\\
D_{\mathrm{pop}}^{(t)}
=&\frac{2}{P(P-1)N}
\sum_{a=1}^{P-1}\sum_{b=a+1}^{P}
\left\|
\mathbf{M}_{a}^{(t)}-\mathbf{M}_{b}^{(t)}
\right\|_0 ,
\end{aligned}
\end{equation} where $P$ denotes the population size and $\mathbf{M}_{a}^{(t)}$ denotes the $a$-th candidate mask at generation $t$. A lower $\bar{H}^{(t)}$ indicates a more deterministic feature-selection policy, while a non-zero $D_{\mathrm{pop}}^{(t)}$ reflects sufficient candidate diversity for continued total-score-driven exploration.

\vspace{-5pt}

As shown in Fig.~\ref{fig_task3_policy}, the policy probabilities gradually concentrate on several informative features and become stable, indicating that the agent forms task-oriented feature preferences through total score feedback. 
Fig.~\ref{fig_task3_agent} further shows that the normalized agent entropy decreases while the population diversity remains non-zero, suggesting that the search becomes more decisive without prematurely collapsing to a single feature subset. 
Overall, these results support the closed-loop feature decision and total score feedback mechanism in the proposed case study.

\vspace{-5pt}

\vspace{-13pt}
\section{Conclusions}\label{sec5} 
\vspace{-2pt}

To address the increasing heterogeneity in future 6G systems, this paper proposes a ChannelAgent-empowered EM space world model, where ChannelAgent serves as the intelligent core for closed-loop sensing, reasoning, decision-making, and feedback update. An agent-driven channel generation case study is instantiated through path loss prediction, in which a hybrid feature selection mechanism combines reinforcement-learning-inspired policy adaptation and evolutionary search. Simulation results demonstrate that the proposed method adaptively derives task-suitable feature subsets and improves path loss prediction performance across different tasks, providing an initial validation of the proposed model. Future work will focus on extending the model to more complex communication tasks and dynamic environments, and enhancing its online decision-making and continual learning capabilities.

\section*{Acknowledgment}

This work is supported by the National Natural Science Foundation of China (No. 62401084, No. 62525101 and No. 62401068), and the Beijing University of Posts and Telecommunications - China Mobile Communications Group Co., Ltd. Joint Institute.

\end{document}